# Real-Time Generation of Leg Animation for Walking-in-Place Techniques

Jingbo Zhao*  Zhetao Wang  Yiqin Peng  Yaojun Wang†

College of Information and Electrical Engineering, China Agricultural University
Beijing, China

## ABSTRACT

Generating forward-backward self-representation leg animation in virtual environments for walking-in-place (WIP) techniques is an underexplored research topic. A challenging aspect of the problem is to find an appropriate mapping from tracked vertical foot motion to natural cyclical movements of real walking. In this work, we present a kinematic approach based on animation rigging to generating real-time leg animation. Our method works by tracking vertical in-place foot movements of a user with a Kinect v2 sensor and mapping tracked foot height to inverse kinematics (IK) targets. These IK targets were aligned with an avatar's feet to guide the virtual feet to perform cyclic walking motions. We conducted a user study to evaluate our approach. Results showed that the proposed method produced compelling forward-backward leg animation during walking. We show that the proposed technique can be easily integrated into existing WIP techniques.

## CCS CONCEPTS

• Human-centered computing—Human computer interaction (HCI)—Interaction paradigms—Virtual reality • Human centered computing—Human computer interaction (HCI)—HCI design and evaluation methods—User studies

## 1 Introduction

Estimating forward walking speed by monitoring in-place lower-body movements for walking-in-place (WIP) techniques has been a well-studied topic [1,2,4,7,12–14,16,18]. However, few studies addressed the problem of visualizing and representing a user's legs in virtual reality (VR) during in-place walking. Animating users' legs during virtual walking enhances one's sense of body ownership and presence in virtual environments [9]. It is desirable and beneficial to add leg animation to WIP techniques to give users a better VR locomotion experience.

In the physical space, we produce forward-backward reciprocating leg motions when we naturally walk. However,

*zhao.jingbo@cau.edu.cn
†wangyaojun@cau.edu.cn

during in-place walking in VR, our feet move upward and downward. We wish to animate our legs in virtual environments such that the virtual leg movement patterns are similar to that of our natural walking.

Generating forward-backward leg animation for WIP techniques generally requires finding a set of mappings between tracked vertical leg movements and forward-backward leg movements in virtual environments. In the current study, we present a kinematic approach to render a user's leg animation while a user is walking in place with a Kinect sensor that tracks a user's foot height. Our main approach is to use IK targets, with a set of derived kinematic equations, to map a user's tracked foot height to cyclic motions of the IK targets. These IK targets are aligned with the feet of a user's avatar in virtual environments. Thus, cyclic motions of the IK targets guide an avatar's feet to produce forward-backward leg animation. Since the derived kinematic equations are dependent on gait phases, we implement gait detection state machines to monitor the gait phases of a user's feet.

The main contribution of the study is that we present a kinematic approach to generating real-time forward-backward leg animation for WIP techniques based on animation rigging using foot tracking data given by the Kinect sensor. Compared to previous methods [5,9,10], the present approach does not need a cycling device and allows leg animation to match physical leg movements as gait phases are directly detected from leg movements. We show that this technique can be easily incorporated into existing WIP techniques that can detect gait phases to generate real-time leg animation during virtual walking. The present paper is an extended version of a previously published abstract [19]. The present paper extends the description of the proposed method with more details and includes the experiment that evaluated the proposed method.

## 2 Related Work

Only a very limited number of studies have attempted to solve this leg animation generation problem for WIP techniques. For example, previous work by Lee et al. [9] tackled this problem by using a 1-D convolutional neural network (CNN) to detect gait phases from the head tracking signals of a VR headset and using the detected gait phases as key-frames to animate pre-defined leg animation. A limitation of this approach is that the predicted gait phases may not be consistent with the actual gait phases of the users.

Thus, the rendered leg animation may not match a user's actual leg movements. Freiwald et al. [5] presented a locomotion interface called VR strider, which consisted of a mini exercise bike, tracking and vibrotactile feedback devices. When a user pedaled the mini exercise bike, tracked circular motions were mapped to the inverse kinematics (IK) targets aligned with the avatar's feet. Leg animation was generated with the guidance of the IK targets. However, this approach requires users to have this modified mini exercise bike as the platform to use this locomotion approach. Park and Jang [10] presented a WIP method that rendered leg animation based on the GUD-WIP technique [16]. A step period was estimated from a user's gait. The estimated step period was subsequently linearly mapped to the angles of the lower limb joints. However, this approach does not allow leg animation to match the motion of users as a step period needs to be estimated first before leg animation can be rendered. To the best knowledge of the authors, our work is the first that aims to solve this problem by using IK targets and kinematic equations driven by tracked foot height to generate compelling leg animation for WIP techniques. Compared to previous work, our method does not require specialized hardware and allows the avatar's leg animation to match the leg motion of a user.

## 3 Methods

### 3.1 Hardware and Software of the VR system

Our experimental VR hardware consisted of a Kinect v2 sensor to track a user's body movements, an Oculus Quest 2 headset to present virtual environments, a computer with an Intel-i5 11400F CPU, 16 GB memory and a Geforce RTX 3060 graphics card with 12 GB graphics memory to host the experimental application software. Implementation of the proposed method and prototyping of the experimental virtual environments were completed using the Unity 2019.4. Animation rigging was supported by the Unity's Animation Rigging version 0.2.7-preview package. The avatar model used in our experiment was selected from the Microsoft Rocketbox Avatar library [14]. The rate of the game logic update was fixed as 30 Hz in sync with the sampling rate of the Kinect v2 sensor.

### 3.2 Mapping Foot Height to Positions of IK Targets

We used the ankle joints tracked by the Kinect v2 sensor as the reference to monitor a user's foot height. The height of the floor plane was estimated by capturing the ankle joints for 3 s when a user stood still during the calibration procedure. The data were then averaged to get a reference height of the floor plane. For both feet, we buffered two data sequences of the *y*-component of the ankle joints with a length of thirty elements using two queues. This corresponded to 1 s of data as the sampling rate of the Kinect v2 sensor is 30 Hz. Vertical speed ($s_{left}$ and $s_{right}$) and velocity ($v_{left}$ and $v_{right}$) were calculated from the *y*-position data for further processing steps. We further smoothed the position data with two low-pass Butterworth filters of 4 Hz to suppress noise. By inspection, it was not necessary to smooth the speed and velocity data. Velocity data were used for gait phase transitions and speed data were used for calculating the average speed between transitions of gait cycles to estimate forward walking speed [1]. The tracked height values of both feet are denoted as $h_{left}$ and $h_{right}$, which are the most recent values from the smoothed data sequences.

We created two two-bone IKs using the recent Animation Rigging package from the Unity and respectively assigned the knee joints and the foot joints to IK hints and IK targets for both legs. According to the documentation of the Animation Rigging package, IK hints are optional, and they control the bend normal of the middle nodes of the two-bone IKs. Thus, IK hints were positioned by trial and error until desirable leg animation could be produced. They were finally placed 0.29 m in front of the avatar's knees in depth. Other transform parameters of IK hints were aligned with the knee joints. The transforms of the IK targets were fully aligned with the avatar's foot joints. This completes the setup for the rest of the calculations and discussion.

Assume that an avatar is producing forward-backward walking animation in place without moving forward. The step length and step height that an avatar can achieve for both feet are denoted by $z_{left}$, $z_{right}$, $y_{left}$ and $y_{right}$. These values were mapped to the IK targets in their local coordinate to guide the forward-backward movement of an avatar's feet. The mapping is symmetrical about the avatar's trunk in depth. Thus, $z_{left}$ and $z_{right}$ can be negative or positive. The horizontal values of the IK targets along the *x*-axis were fixed.

For analyses of the IK targets' movements, we expand our discussion into three cases: a user's first step, consecutive walking steps and the stopping step. In addition, we consider three gait phases: the double support phase, the ascending phase, and the descending phase, for updating the leg animation. Other gait phases, including the initial swing, the mid-swing, and the terminal swing [11], are transitions of the gait detection state machine triggered by velocity and position conditions (see Figure 3 below). Leg animation is not updated in these transitions. The kinematic equations derived in this study came from our inspection on recorded treadmill foot position data and direct observation on human walking. Since a walking cycle consists of three different gait phases and it is also necessary to process the first step, consecutive steps and the stopping step, using a single equation to fit recorded foot position data obtained from treadmill walking is not feasible. Thus equations 1-4 have been derived as piecewise functions to produce foot trajectories in depth, based on a user's tracked foot height.

#### 3.2.1 First Step

Initially, the avatar stands still in an initial double support phase with both feet centered on the origin in their local coordinate. The step length and step height for both feet are zeros. The user also starts with a double support phase (Figure 1a). Assume that the user lifts their left foot first (Figure 1b) and ascends to the mid-swing phase. This requires us to map the left stride of the avatar forward in depth and the right stride backward during the left foot's ascending phase (denoted by ↑ in the equations for compact representation). Once we detect the mid-swing, the left foot enters the descending phase (denoted by ↓) and plants on the ground subsequently. We describe this process using two piece-wise functions for ascending and descending phases of both feet for an avatar:

$$z_{left} = \begin{cases} -0.25 z_{max} \cos(2\pi h_{left}/H) + 0.25 z_{max}, \uparrow \\ 0.25 z_{max} \cos(2\pi h_{left}/H) + 0.25 z_{max}, \downarrow \end{cases} \quad (1)$$

$$z_{right} = \begin{cases} 0.25 z_{max} \cos(2\pi h_{left}/H) - 0.25 z_{max}, \uparrow \\ -0.25 z_{max} \cos(2\pi h_{left}/H) - 0.25 z_{max}, \downarrow \end{cases} \quad (2)$$

where $H$ is a scale factor that enables the function to complete a gait cycle and $H = 4 y_{max} = 1.2\ m$. $z_{max}$ ($z_{max} = 0.6\ m$) is the maximum step length that an avatar can produce and $y_{max}$ ($y_{max} = 0.3\ m$) is the maximum step height of a user. See Figure 2a and Figure 2b for the plotted curves of equations 1 and 2.

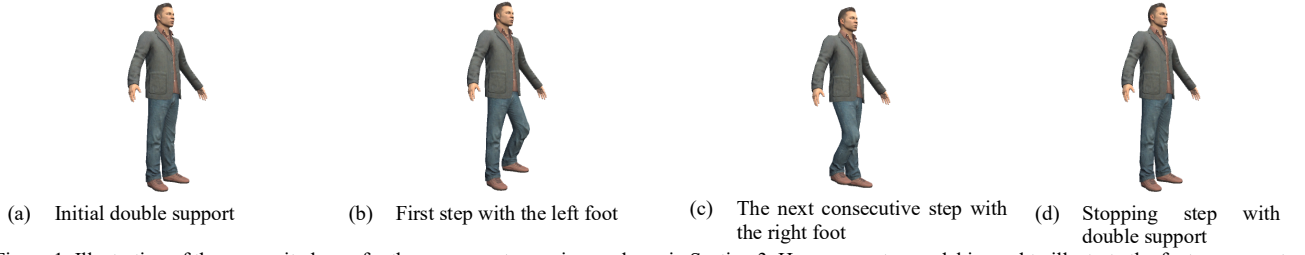

Figure 1: Illustration of the user gait phases for the movement mapping analyses in Section 3. Here, an avatar model is used to illustrate the foot movements of the user during in-place walking. The present model was selected from the Microsoft Rocketbox Avatar Library.

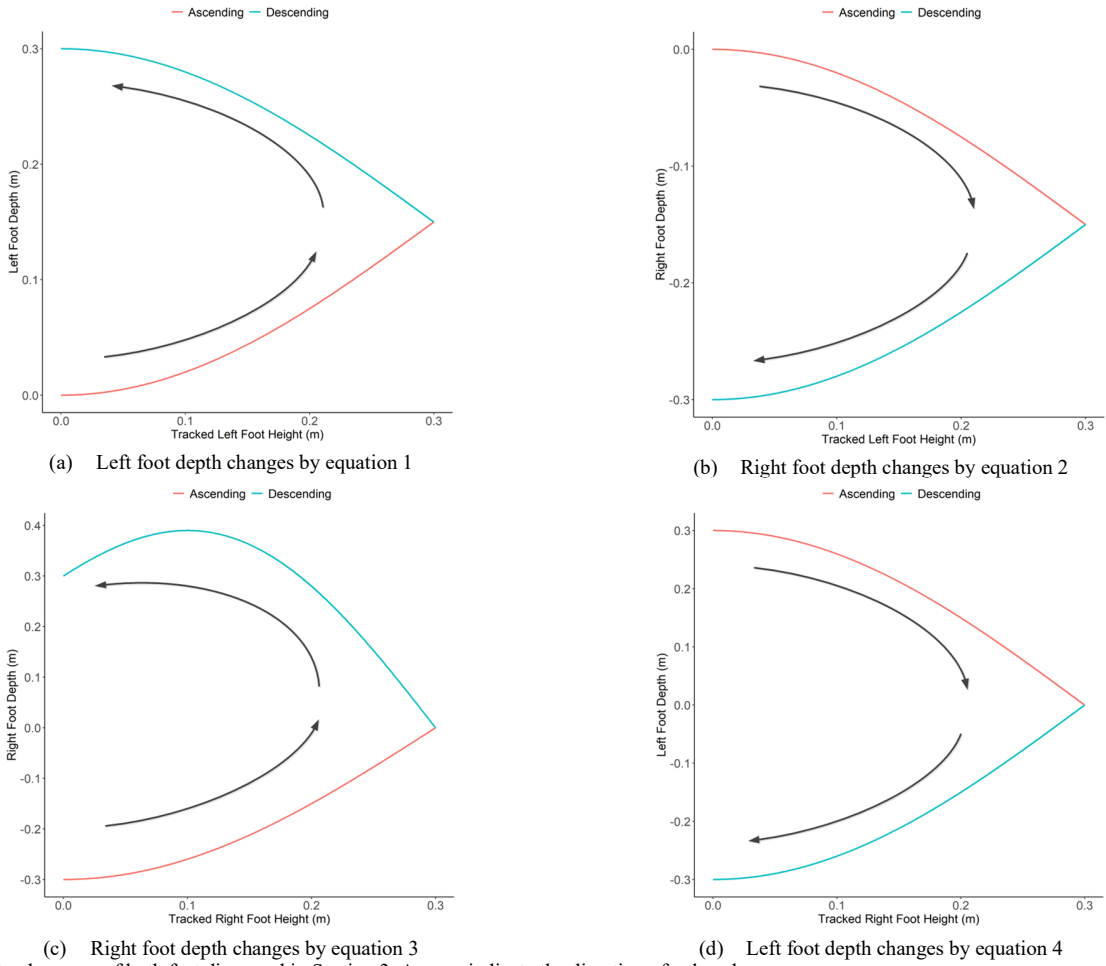

Figure 2: Depth curves of both feet discussed in Section 3. Arrows indicate the direction of value changes.

When the avatar finishes the very first step, the avatar goes into double support with the left stride and right stride symmetrical about the trunk. From this gait phase, we discuss the animation mapping of consecutive walking steps.

### 3.2.2 Consecutive Walking Steps

Subsequently, the user supports the trunk using the left foot. The right foot (opposite foot) is lifted (Figure 1c) and ascends to the mid-swing and then descends to foot support again. Since the mapped avatar leg motion is symmetrical about the trunk, we describe this process using another piece-wise function:

$$z_{right} = \begin{cases} -0.5z_{max}cos(2\pi h_{right}/H), \uparrow \\ 0.5z_{max}cos(2\pi h_{right}/H) + 0.25z_{max}sin(4\pi h_{right}/H), \downarrow \end{cases} \quad (3)$$

In equation 3, a correction term $0.25z_{max}sin(4\pi h_{right}/H)$ is added to enable the leading foot to strike the ground more naturally during the terminal swing.

When the right foot of the avatars is mapped to move forward, the left foot needs to be retracted and trailed behind the trunk, and this can be described by:

$$z_{left} = \begin{cases} 0.5z_{max}cos(2\pi h_{right}/H), \uparrow \\ -0.5z_{max}cos(2\pi h_{right}/H), \downarrow \end{cases} \quad (4)$$

Note that $h_{right}$ is used in equations 3 and 4 as the animation of this second step is controlled by the user's right foot. The equations 3 and 4 hold for the next consecutive step (i.e. left foot) by swapping the symbols representing the parameters of a user's left and right feet. See Figure 2c and Figure 2d for the plotted curves of equations 3 and 4.

The equations described above were implemented using two gait detection state machines (discussed in Section 3.4) to respectively monitor the gait phases of a user's left foot and right foot and control their respective IK targets. Motions of IK targets are smoothed using the lerp() function from the Unity engine.

### 3.2.3 Stopping Step

We consecutively monitor the elapsed time $t$ in the double support phase of the user (Figure 1d). Whenever $t$ is larger than an empirical value of 1.5 s. We consider that the user intends to stop and we set $z_{left}$ and $z_{right}$ to zeros using the lerp() function to move the avatar's legs to the initial standing state with double support.

## 3.3 Manipulation of an Avatar's Vertical Position

When we naturally walk on the ground, our body moves up and down. If our left and right strides reach the maximum position in depth (double support), our height will be much lower compared to normal standing. For avatar walking animation, such height change during walking also needs to be considered. We derive an equation that relates the avatar's vertical position to the step length of the trailing foot (left foot in our analysis) of a user:

$$\Delta h = l_{leg} - \sqrt{l_{leg}^2 - z_{left}^2} \quad (5)$$

where $\Delta h$ is the amount of the adjustment for the avatar's vertical position, $l_{leg}$ the leg length of the avatar ($l_{leg}$ = 0.8 m in our selected model for analysis), manually determined from the avatar model parameters, and $z_{left}$ the z-position of the trailing foot assume that the right foot is the leading foot.

The vertical position of the avatar during walking is updated as:

$$y_{curr} = y_{init} - \Delta h \quad (6)$$

where $y_{init}$ is the initial vertical position before the avatar starts walking, and $y_{curr}$ the updated vertical position value assigned to the avatar.

The vertical position of the avatar when they stop is updated as:

$$y_{curr} = y_{init} \quad (7)$$

Finally, for both cases: the first step and consecutive walking steps, mapping the vertical foot height $y_{left}$ and $y_{right}$ for the avatar is trivial. Assume that the user's left foot ascends and then descends in motion and the right foot supports the trunk during the process, the mapping can be derived as:

$$\begin{cases} y_{left} = 0.5h_{left} + \Delta h, \uparrow \ and \ \downarrow \\ y_{right} = \Delta h, \uparrow \ and \ \downarrow \end{cases} \quad (8)$$

for both ascending and descending phases. We comment that a scale factor of 0.5 is an empirical value. Further investigation is required to determine if a scale factor of 0.5 or other value is appropriate for vertical mapping. Parameter $\Delta h$ adjusts the vertical position considering the height changes of the avatar body during walking. For equations 5-8, the lerp() function was also used to smooth motion.

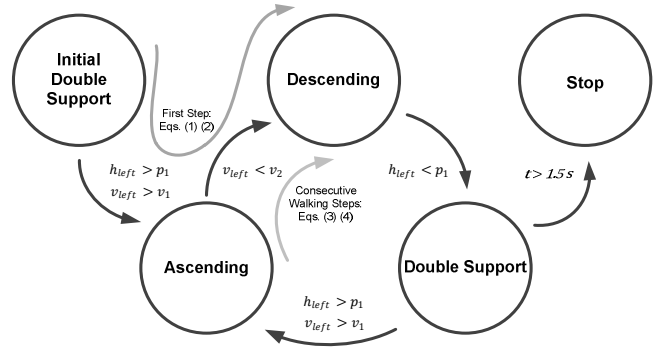

Figure 3: State diagram using the left foot as an example that shows transitions between different gait phases starting with an initial double support and subsequently followed by ascending, descending and double support. When the time $t$ is larger than 1.5 s, we determine that the user intends to stop. Black arrows show the direction of transitions between states with the associated conditions ($p_1 = 0.1 \ m, v_1 = 0.2 \ m/s$ and $v_2 = 0.6 \ m/s$). Grey arrows show the equations involved with their associated states. To apply this diagram to the right foot, it is necessary to swap the symbols in both the diagram and equations (i.e. $h_{left} \rightarrow h_{right}$ and $v_{left} \rightarrow v_{right}$, etc.).

### 3.4 Gait Phases Detection and Walking Speed Estimation

We selected the WIP technique by Bruno et al. [1] and integrated our animation method into this technique to implement our method. To monitor the gait phases, we implemented two state machines to monitor the gait phases of two feet of a user. The original implementation of the Bruno et al.'s work [1] was based on an optical tracking system with higher precision and a faster sampling rate than the Kinect v2 sensor. We empirically adjusted the parameters of the state machines to enable them to monitor gait phases. Walking speed estimation and the stopping criterion for producing zero speed were based on the work by Bruno et al. [2]. The readers are referred to the speed estimation model in their work for walking speed estimation and the criterion for determining walking stop. As in their work, we only adapted the walking speed whenever the mid-swing and the terminal swing were detected. The estimated walking speed was used to set the forward moving speed of the avatar along the $z$-axis. The required time threshold to stop walking can be calculated using the stop time estimation equation in Bruno et al.'s work [2]. However, the determination of stopping to move the avatar's leg back to the neutral position in double support required an empirical constant time threshold of 1.5 s to avoid false positive detection of walking stop.

In Figure 3, we present a state diagram to summarize the relationship between the gait phases and the derived equations assuming the left foot is lifted first. Applying the state diagram to the right (opposite) foot only requires changing the parameter symbols from the left foot to the right foot (i.e. $h_{left} \rightarrow h_{right}$ and $v_{left} \rightarrow v_{right}$, etc.). Values of position and velocity thresholds to determine state transition are also given in Figure 3.

While in the current study we selected the WIP technique by Bruno et al. [1,2] and integrated our leg animation approach into this WIP technique to test our method, our method also can be incorporated into other WIP techniques, such as the GUD-WIP [16], that have gait phases detection mechanisms. Since our method only relies on users' tracked foot height to generate foot trajectories, it can also work with other sensor systems that can track users' foot height.

### 3.5 Key-frame Leg Animation for Experimental Comparison

We designed two key-frame leg animations for walking and standing postures using the animation rigging tool to compare our proposed method. The same gait detection state machine (described in Section 3.4) was used to detect whether a user's feet are moving or not. The real-time IK targets trajectory generation equations were removed. Whenever we detected that a user's feet ascended or descended, the walking animation was played. When the user stopped, the standing animation was played. The key-frame animation was not synchronized with leg movements. Leg animation was played when ascending and descending phases were detected by state machines.

## 4 Experiment

### 4.1 Introduction

The goal of the experiment was to evaluate whether our proposed method can generate compelling leg animation when users walk in virtual environments and understand difference between two animation conditions: the key-frame animation and the real-time animation.

### 4.2 Participants

We invited 12 volunteer undergraduate students (8 males and 4 females, age: 20-22, height: 155 cm-183 cm) to participate in our experiment. All had normal or corrected-to-normal vision. Informed consents were provided to participants and completed before experiments.

### 4.3 Design

We designed a 3D environment that included a plane with a repeated grass texture representing the ground and a walking path with concrete squares texture for virtual walking. The trees placed along the walking path were selected from the Free SpeedTrees Package from the Unity Asset Store. An experimental session was designed to have two experimental conditions (blocks): (a) real-time leg animation generated using the method described in Section 3; (b) replay of key-frame leg animation generated using pre-defined key frames with the animation rigging tool.

For each block, there were one practice trial for participants to get familiar with the task and two experimental trials. A questionnaire was adapted from [8] based on a 7-point Likert scale (from strongly disagree to strongly agree) to ask participants to subjectively evaluate their leg animation. The questionnaire included the following statements: (1) Leg animation was smooth; (2) Leg animation was natural; (3) Leg animation was matched to my leg motion; (4) Leg animation was the same as I intended. For short notation, we refer to these factors as smoothness, naturalness, match and intention. Whenever a participant finished a complete block, they were asked to complete the questionnaire to assess their leg animation during walking. We controlled for order effects by dividing participants into two groups and counter-balancing the order of the blocks. We created a rigid body for the avatar used in the experiment and disabled its gravity option in the Unity. The estimated forward walking speed was assigned to the $z$-axis of the rigid body of the avatar to enable it to move forward. We limited the maximum forward walking speed to 3.5 m/s during experiments. Each experimental trial required a participant to walk 50 m to complete the trial. A collision volume is placed 50 m ahead of the avatar to determine the completion of the trial. The Oculus OVR camera rig was placed at the eye position of the avatar model for participants to view the virtual environment and their virtual body.

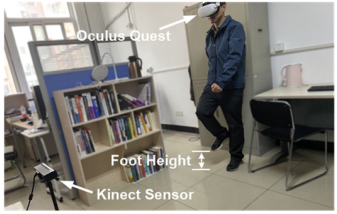

Figure 4: Experimental setup. A user wearing an Oculus Quest 2 headset was walking in place, captured by a Kinect sensor.

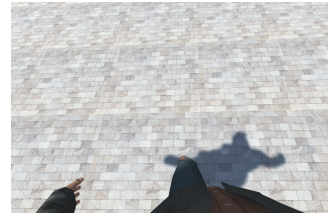

Figure 5: First-person view during walking. Participants were asked to look down and observe their virtual legs to evaluate animation results.

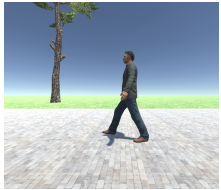 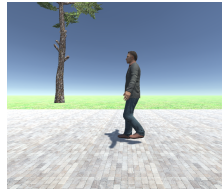 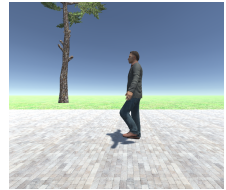 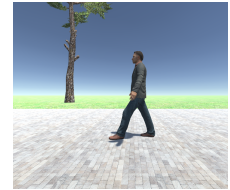

(a) Double Support     (b) Initial Swing     (c) Mid Swing     (d) Terminal Swing

Figure 6: An exemplar sequence of screenshots of an avatar walking captured from a side view while a user was walking in place, with foot height tracked by a Kinect sensor.

### 4.4 Procedure

We mounted the Kinect v2 sensor on a tripod approximately 60 cm high from the ground, with the camera view capturing a participant (see Figure 4 for the experimental setup). During an experimental session, we first introduced the experimental task to a participant and provided an informed consent for the participant to sign. The participant was then asked to stand approximately 1.8 m in front of the Kinect sensor wearing the Quest 2 headset. The Quest 2 headset was wirelessly connected to the host computer using the Airlink feature through a TP-Link Wi-Fi 6 router. Participants were asked to look down and observe their virtual legs during walking. The researcher then asked the participant to level their heads and reset the view so that the participant's view could be aligned to the eye position of the avatar. Then, a 3-s calibration was conducted, capturing the participant's foot position to determine the reference height of the floor plane for calculating foot height. After calibration was done and when the participant was ready, the researcher pressed the start button to initiate the trial with a 3-s countdown timer. When this timer was up and a "start" prompt was shown, the participant started walking until they reached the collision volume to complete the trial. Participants were asked to fill the questionnaire once they finished a complete block. After a short rest of 5 min, participants were introduced to the second block, and they completed the block following the same procedure discussed above.

Figure 5 shows the first-person view of a user while walking in place. Figure 6 shows an exemplar sequence of screenshots showing an avatar walking, controlled by a user's foot height, from a side view. These images show four different gait phases and present the concept of the animated result.

### 4.5 Results

In Figure 7, we present the results of the analyses on subjective factors. A Wilcoxon rank sum test on the factor "Match" in the animation evaluation questionnaire revealed a significant effect ($p = 0.012$), which showed that the real-time animation condition clearly matched participants' leg movements better than the key-frame animation condition. However, Wilcoxon rank sum tests on other factors did not reveal significant effects: smoothness ($p = 0.98$), naturalness ($p = 0.86$) and intention ($p = 0.51$). We speculated that the first-person view in VR with legs was a new experience to participants. Considering that the ratings toward both interfaces were high and none had previous experience playing VR games that render leg animation, both animation types were appealing to participants. Thus, high ratings were given to both animation conditions. On the other hand, few VR games present leg animation for self-avatars in virtual environments, so it was difficult for participants and the authors to compare both types of animation in the experiments to existing VR titles or previous VR experience. In terms of statistical power, our design was similar in size to most within-subject studies in human-computer interaction

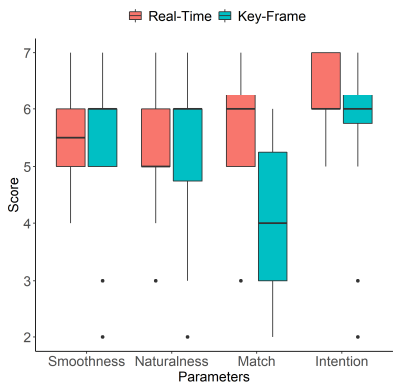

Figure 7: Distribution of subjective ratings based on the animation evaluation questionnaire. Line in a box denotes median, top and bottom of a box denote upper quartile and lower quartile and whiskers denote maximum and minimum.

[3] and was sufficient to demonstrate differences between the types of animation. It is certainly possible that small differences may be found with a larger number of participants. But we believe we had sufficient power to identify meaningful differences as the present technique provides better matching between users' physical leg movements and the leg animation. When there is a requirement to implement leg animation in a VR system, it should be reasonable to use this new technique as it is easy to implement and enables better matching between physical leg motions and rendered leg animation.

## 5 Discussion

One limitation with the proposed approach is that the maximum step length and height are fixed, so it is impossible to generate leg animation with variable step length and height based on one's tracked foot height. Future work should focus on deriving more flexible mappings to enable leg animation to have variable step length and step height. The Quest Pro headset has been recently released by Meta and it has been shown that the device is capable of tracking motions of a user's body parts without using additional hardware or trackers. This makes it possible to remove the use of an optical sensor, such as the Kinect sensor, to track body motions for the proposed leg animation technique. The proposed method can be adapted to this new device, making it accessible to general VR users, who often only possess a VR headset.

Further study of the proposed method includes comparing it to methods proposed by other researchers [9,10], investigating its utility in more general virtual locomotion cases and comparing the effects of the present method and other methods on gait variability. In addition, Waltemate et al. [15] presented a method for generating personalized avatars for users in virtual environments. It will be interesting to leverage the method to create personalized avatars for individual users and investigate if such manipulation affects user experience on animation, body ownership and presence when animating users' virtual legs during walking. Research by Gonzalez-Franco et al. [6] has shown that users tend to follow the movements of the avatars and Willaert [17] reported that users' gait can be modulated by avatars. While setting up the key-frame leg animation experimental condition, we also observed that users may follow the unsynchronized leg motions under this condition. Further research is needed to study the influence of the self-avatar follow effect on gait parameters, presence, and body ownership when leg animation is not in sync with a user's leg motion.

## 6 Conclusion

In this work, we presented a kinematic approach to generating forward-backward leg animation based on animation rigging, with foot height tracked by a Kinect v2 sensor. A set of kinematic equations were proposed to guide the legs of an avatar to generate leg animation. Our user study shows that the present method produces compelling leg animation and better matching between a user's physical leg and an avatar's leg than the keyframe based animation. Thus, this method is promising to be applied to practical VR WIP locomotion applications or experiences, where rendering of leg animation is needed.